\documentclass[letter]{aa} % for the letters
\usepackage{amssymb}
\usepackage{amsmath}

\usepackage{graphicx}
\usepackage{graphicx}
\usepackage{epsfig}
\usepackage{graphics}

\begin{document}

\title{On the nature of the AGILE galactic transient sources}
\author{Gustavo E. Romero\inst{1,2,}\thanks{Member of CONICET, Argentina} \and Gabriela S. Vila\inst{1,}\thanks{Fellow of CONICET, Argentina}}
  
\institute{Instituto Argentino de Radioastronom\'{\i}a (IAR), CCT La Plata  (CONICET), C.C.5, (1894) Villa Elisa, Buenos Aires, Argentina \and Facultad de Ciencias Astron\'omicas y Geof\'{\i}sicas, Universidad Nacional de La Plata, Paseo del Bosque s/n, 1900, La Plata, Argentina}

\offprints{G. E. Romero \\ \email{romero@iar-conicet.gov.ar}}

\titlerunning{AGILE sources}

\authorrunning{G.E. Romero and G.S. Vila}

\abstract
{The Italian gamma-ray satellite AGILE has recently reported the detection of some variable high-energy sources likely of galactic origin. These sources do not have any obvious counterpart at lower energies.}
{We propose that these sources are produced in proton-dominated jets of galactic microquasars.}
{We develop a model for microquasar jets that takes into account both primary leptons and protons and all relevant radiative processes, including secondary particle emission and gamma-ray attenuation due to pair creation in the jet.}
{We obtain spectral energy distributions that correspond to what is observed by AGILE, with most of the power concentrated between 100 MeV and 10 GeV and reaching luminosities of $10^{34-35}$ erg s$^{-1}$. We make detailed spectral predictions that can be tested by the Fermi gamma-ray telescope in the immediate future.}
{We conclude that hadronic jets in galactic accreting sources can be responsible for the variable unidentified gamma-ray sources detected by AGILE.}
 
\keywords{X-rays: binaries - gamma rays: theory - radiation mechanisms: non-thermal } 
 
\maketitle
 
\section{Introduction}

The Italian gamma-ray satellite AGILE (Astro-rivelatore Gamma a Immagini Leggero) has detected several non-identified variable sources likely of galactic origin. These sources include the strong source AGL J2021+4029 located in the Cygnus region, with the center of gravity of the error box at $l=78.01$ deg and $b=2.19$ deg (Longo et al. 2008), the variable source in the Musca region (error box centered at $l=312.2$ deg and $b=-0.3$ deg, Pittori et al. 2008), and the high galactic latitude transient AGL J0229+2054, with the error box centered at $l=151.7$ deg and $b=-36.4$ deg (Bullgarelli et al. 2008). The Cygnus source showed some significant re-brightening after its discovery (Giuliani et al 2008, Chen et al. 2008). Simultaneous X-ray observations with Super-AGILE did not show any counterpart in the 20-60 keV band. A steady and weak source was detected within the large error box by Swift/BAT (15-55 keV, see Ajello et al. 2008), but there is no clear relationship. Additional X-ray observations have been performed with XMM-Newton, without adding new clues (Pandel et al. 2008). Radio observations with the Very Large Array (VLA) have shown no clear counterpart (Cheung 2008).   

Concerning the source in the Musca region, a potential archival X-ray counterpart has been claimed on the basis of BeppoSAX observations dating from January 2001 (Orlandini et al. 2008). The high-latitude transient, on the other hand, might be a halo galactic source or a blazar (the AGN 1ES 0229+200 is at 61.3 arcmin from Gamma-Ray Imaging Detector error box centroid). 

The fact that these sources are highly variable implies that the high-energy radiation should be produced in a compact region. The absence of detection with Super-AGILE means that the ratio of gamma-ray to X-ray luminosities $L_{\gamma}/L_{X}$ should be $>>1$. These characteristics recall those of the population of variable EGRET sources found in the Galactic Plane and in the Halo (Romero 2001; Grenier 2001, 2004; Nolan et al. 2003). Actually, the AGILE detections in the Cygnus region and the Musca region partially overlap with the location error box of the sources \mbox{3EG J2020+4017} and \mbox{3EG J1410-6147}, respectively. It has been proposed that the unidentified variable gamma-ray sources sources at MeV-GeV energies might be high-mass microquasars with the emission dominated by inverse Compton up-scattering of UV stellar photons from the hot donor star (Kaufman-Bernad\'o, Romero \& Mirabel 2002, Bosch-Ramon, Romero \& Paredes 2005). On the Galactic Plane, the donor star could be strongly obscured, rendering difficult its detection. These models, however, predict a significant production of X-rays, something that is at odds with the new AGILE and Super-Agile observations. Grenier et al. (2005) proposed that the variable high-latitude unidentified sources might be old, low-mass microquasars expelled long ago from the Galactic Plane or from globular clusters (see Mirabel et al. 2001). They also showed that external Compton models cannot account for the energetics required by the sources.  Romero \& Vila (2008) and Vila \& Romero (2008) showed that ``proton" microquasars with low-mass donor stars might explain the halo EGRET sources through proton synchrotron radiation and photo-meson production.

In this {\em Letter} we propose that microquasars with proton dominated jets can produce spectral energy distributions that satisfy all constraints imposed by AGILE observations and we make some predictions that can be used to test our proposal with the GLAST-Fermi satellite, and ground-based Cherenkov telescope arrays.      

\section{The model}

The model we are going to present is based on the galactic jet model described by Romero \& Vila (2008), but it incorporates several refinements. The jet is assumed to carry most of the accretion power, in accordance with the dissipationless disk model of Bogovalov \& Kelner (2005, 2008).

\begin{figure*}[!t]
  \centering
  \includegraphics[trim=10 10 0 10,clip, width=0.42\textwidth, keepaspectratio]{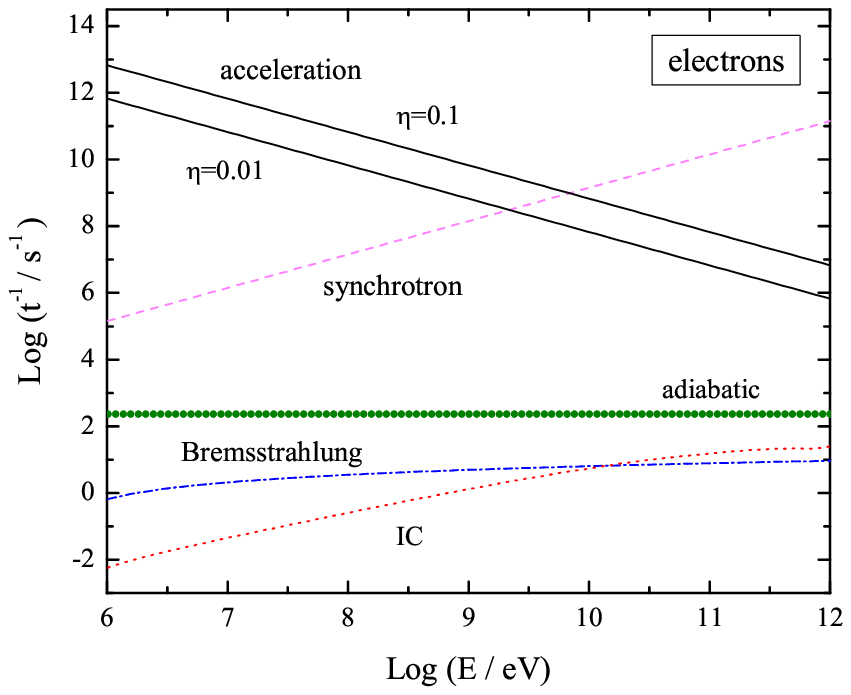}
  \includegraphics[trim=10 10 0 10,clip, width=0.42\textwidth, keepaspectratio]{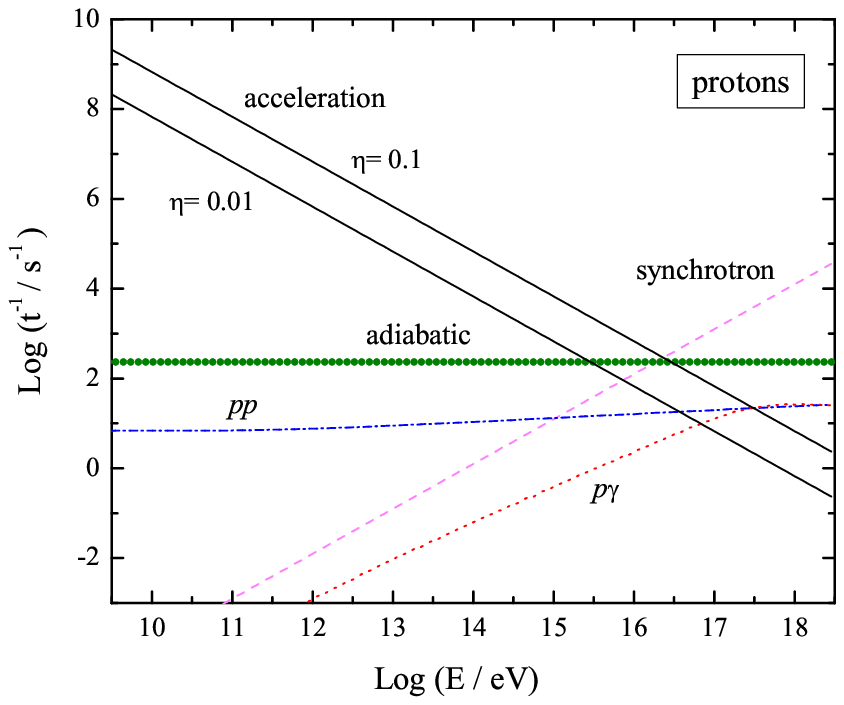} 
  \includegraphics[trim=10 10 0 10,clip, width=0.42\textwidth, keepaspectratio]{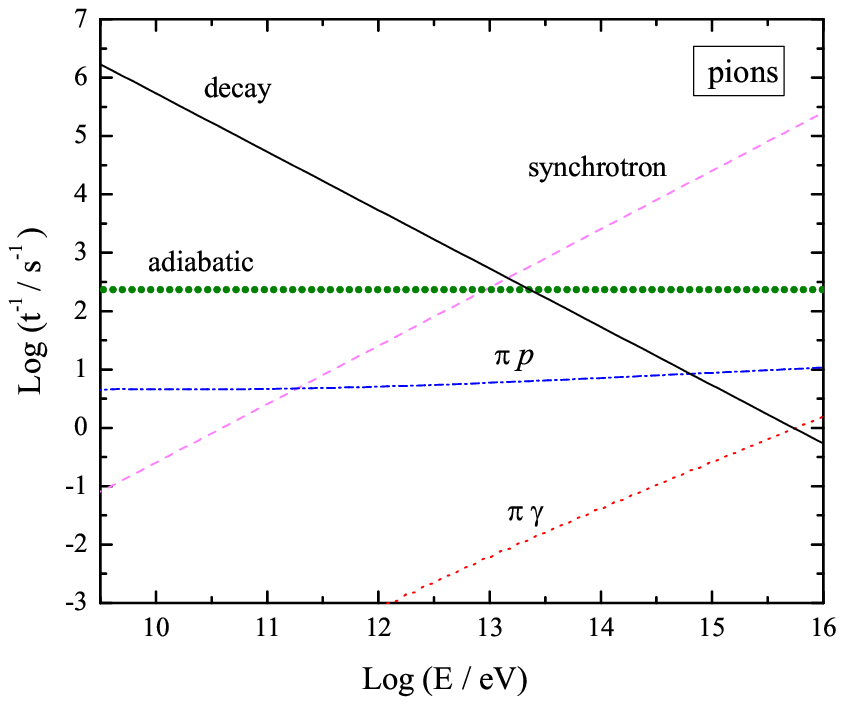} 
  \includegraphics[trim=10 10 0 10,clip, width=0.42\textwidth, keepaspectratio]{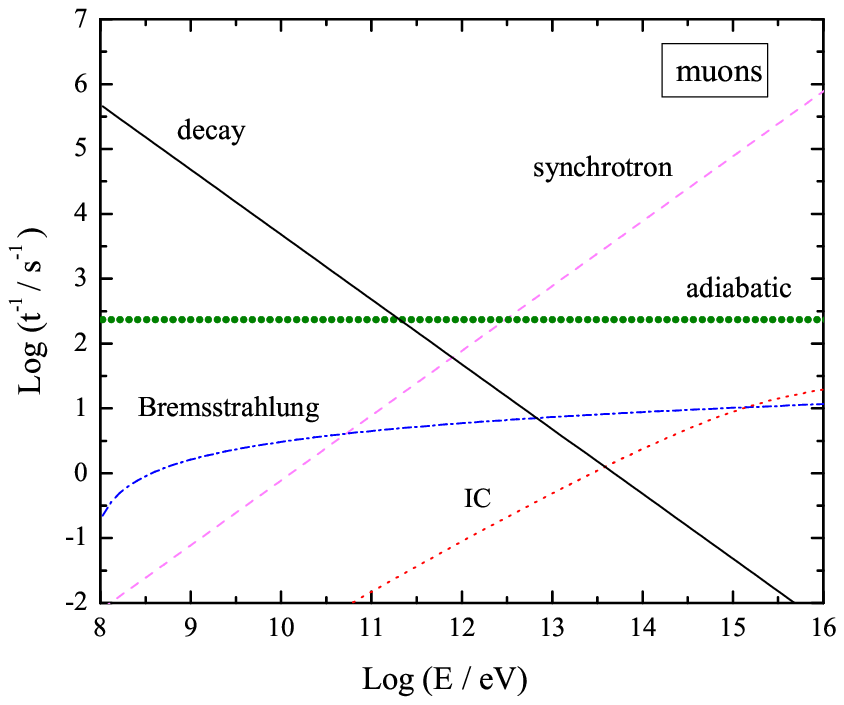}    
  \caption{Acceleration and cooling rates at the base of the jet for primary protons and electrons, and secondary pions and muons, calculated for representative values of the model parameters (proton-to-lepton energy ratio $a=1000$, and primary injection spectral index $\alpha=1.5$). The value of the magnetic field is $B\sim10^7$ G, and the acceleration efficiency parameter $\eta$ is indicated.} 
  \label{fig:cool}
\end{figure*}

A small fraction of the jet power is transformed into relativistic particles in a `one-zone' acceleration region located close to the compact object (Khangulyan, Aharonian \& Bosch-Ramon 2008). In this region, the magnetic field (obtained by equipartition) is extremely high (of the order of $10^{7}$ G). This produces the immediate cooling of primary electrons and significant cooling of protons. In Fig. \ref{fig:cool} we show the cooling rates for both primary electrons and protons in the acceleration region, as well as the cooling and decaying rate of secondary muons and pions. The maximum energies of the primary particles are obtained equating the cooling rates to the acceleration rate, which is assumed to be produced by mildly relativistic shocks at the base of the jet through a first-order Fermi mechanism. In this work we adopt an acceleration efficiency of the order of 10 \%, which rougly corresponds to a shock velocity of $0.3c$ and a mean free path similar to the Larmor radius. We see that electrons, even for such a high acceleration efficiency, reach only GeV energies, whereas protons can attain much higher energies, well into the PeV band. 

Our model takes into account: 1) synchrotron emission from both types (electrons and protons) of primary particles, as well as emission from secondary leptons (electrons, positrons, and muons) and hadrons (charged pions), 2) inverse Compton (IC) emission from all leptons in the strong total radiation field in the emitting region, 3) photo-pair and photo-meson production by both protons and pions, 4) inelastic collisions between relativistic protons in the jet and the cold material that forms most of the same outflow (see Bosch-Ramon, Romero \& Paredes 2006, for details of the physics of a cold matter dominated jet), 5) relativistic Bremsstrahlung from electrons and muons,  and 6) internal photon absorption in the local radiation fields (calculated as in Aharonian, Khangulyan, \& Costamante 2008), 7) re-injection of secondary pairs, which is negligible due to the low opacity of the proton-dominated jet (see Romero \& Vila 2008 for a discussion). The effect of losses of muons and pions in a strong magnetic field on the resulting gamma-ray and secondary lepton spectra has been recently discussed by Reynoso \& Romero (2008) and we adopt here their treatment. For the other processes we follow the classical formulae (Blumenthal \& Gould 1970), the photo-meson production treatment already used by Romero \& Vila (2008), and the expressions given by Kelner, Aharonian, \& Bugayov (2006) for $pp$ interactions (this formalism is only valid for proton energies above 0.1 TeV; at lower energies we use the standard treatment).

The ratio of relativistic protons to electrons $a$ in the jet is unknown and is used as the basic free parameter in our model. We have calculated a number of models for different acceleration efficiencies, and a set of standard parameters: black hole mass $\sim 10$ $M_{\odot}$, accretion rate $\sim 10^{37}$ erg s$^{-1}$ (i.e., $\sim 10^{-2}$ of the Eddington luminosity), injection $\propto E^{-1.5}$, viewing angle $30^\circ$, jet bulk Lorentz factor $\Gamma=1.5$, and location of the acceleration zone at $z_0\approx10^8$ cm (see Romero \& Vila 2008). 

The transport equation is solved for all types of particles. At steady state it reads:

\begin{equation}
	\frac{\partial}{\partial E}\left[\left.\frac{dE}{dt}\right|_{\rm{loss}}N(E,z)\right]+\frac{N(E,z)}{t_{\rm{esc}}}=Q(E).
	\label{transpeq1}
\end{equation} 
where $t_{\rm{esc}}$ is the particle escape time from the acceleration region ($t_{\rm{esc}}\approx \Delta z/v_{\rm{jet}}$) and $Q$ is the injection function that can be normalized in accordance with the energy budget of relativistic particles through:

\begin{equation}
 L_{e,p}=\int_{V}\mathrm{d}^3r\int_{E_{e,p}^{\rm{min}}}^{E_{e,p}^{\rm{max}}\left(z\right)}\mathrm{d}E_{e,p}\,E_{e,p}\,Q_{e,p}(E_{e,p}).
\label{norminj}
\end{equation}    

The minimum kinetic energy is taken to be of the order of the rest mass energy of the corresponding particle. The equation for pions and muons has an additional term  $N(E,z)/t_{\rm{dec}}$, that takes into account the decay of the particles on a timescale $t_{\rm{dec}}$ that depends on the energy in the lab system. 
  
In general, variability can be obtained introducing a variable injection $Q(E, t)$.  Once the particle distributions are known, the radiative output can be calculated as mentioned above. The internal attenuation results from photon annihilation: $\gamma + \gamma \rightarrow e^{-}+ e^{+}$. The opacity for a gamma ray of energy $E_{\gamma}$ is:

\begin{eqnarray}
	\tau(E_{\gamma})=\frac{1}{2}\int_{l}\,\int^{\epsilon_{\rm max}}_{\epsilon_{\rm th}}\int^{u_{\rm max}}_{-1} (1-u)&& \sigma(E_{\gamma}, \epsilon, u)\nonumber\\ &&\times n(\epsilon, z)\; \mathrm{d}u \; \mathrm{d}\epsilon \;\mathrm{d}l,
\end{eqnarray}

\noindent where $n(\epsilon, z)$ is the photon number density at energy $\epsilon$ and location $z$, $u=cos \vartheta$, $\vartheta$ is the angle between the momenta of the colliding photons, $l$ is the $\gamma$-ray photon path, and the cross section for the interaction is given by (e.g. Levinson 2006):
\begin{eqnarray}
	\sigma_{\gamma\gamma}(E_\gamma,\,\epsilon,&&\vartheta)=\frac{3}{16}\sigma_{\rm T}(1-\beta^2)\nonumber\\&&\times\left[\left(3-\beta^4\right)\ln\left(\frac{1+\beta}{1-\beta}\right)-2\beta\left(2-\beta^2\right)\right].
\end{eqnarray}
Here, $\beta$ is the speed of the electron/positron in the center of momentum frame, i.e.,
\begin{equation}
(1-\beta^2)=\frac{2m_e^2c^4}{(1-u)E_\gamma\epsilon}; \qquad 0\leq\beta<1.	
\end{equation}
The threshold energy $\epsilon_{\rm th}$ is defined by $\beta=1$ with $\vartheta=0$.

\section{Results} 

\begin{figure*}[!t]
  \centering
  \includegraphics[trim=10 15 0 10,clip, width=0.45\textwidth, keepaspectratio]{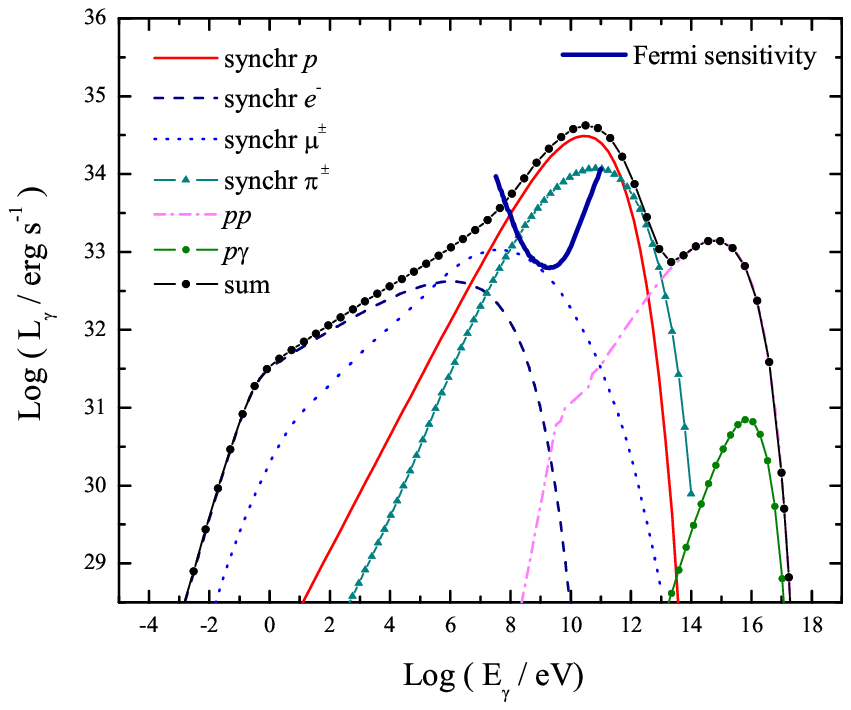}
  \includegraphics[trim=10 15 0 10,clip, width=0.45\textwidth, keepaspectratio]{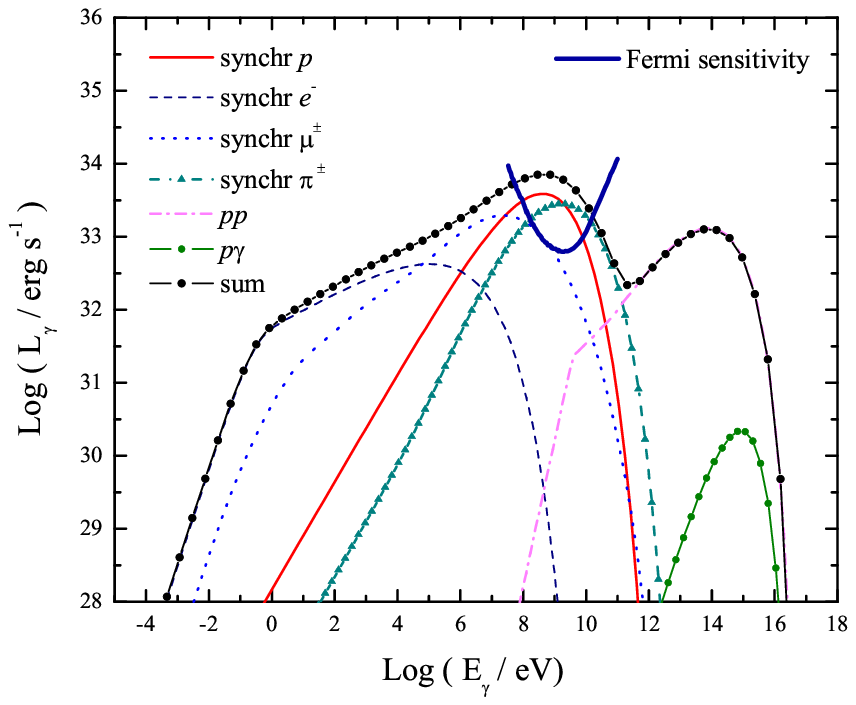} 
  \caption{Spectral energy distributions of a proton-dominated microquasar ($a=1000$). Each panel corresponds to a different acceleration efficiency ($\eta=0.1$ on the left, $\eta=0.01$ on the right). Fermi point source sensitivity is indicated by a thick blue line.}
  \label{SED}
\end{figure*}

In Fig. \ref{SED} we show the spectral energy distribution (SED) calculated for a proton dominated ($a=1000$) jet, with a total power of $10^{37}$ erg s$^{-1}$.  We show the different contributions from all significant cooling processes for all primary and secondary particles, in the case of two different acceleration efficiencies $\eta $ (0.1 and 0.01, from left to right, respectively). Bremsstrahlung and IC losses are negligible, since synchrotron radiation is the dominat cooling channel for leptons. The different acceleration efficiencies not only affect the maximum energy of the radiation but modify the whole shape of the SED, since the dominant synchrotron losses follow the square of the energy and depend significantly on the mass of the particles, resulting in a global photon redistribution. In both cases, however, the peak of the SED is determined by proton synchrotron radiation, followed by pion-synchrotron emission. For the higher efficiency the synchrotron peak is sharper, reaching almost $10^{35}$ erg s$^{-1}$. In the case of a lower efficiency, the peak is slightly above $10^{34}$ erg s$^{-1}$. In the first case most of the emission is concentrated in the range $10^{8}-10^{12}$ eV, whereas in the second it is between $10^{8}$ and $10^{10}$ eV, with a soft slope beyond $10^{9}$ eV. In both cases there is a high ratio $L_{\gamma}/L_{X}$, in accordance with what is inferred from AGILE observations. Soft X-rays, due mainly to electron synchrotron radiation, are at the level of $10^{32}$ erg s$^{-1}$. The hard X-ray component is dominated by muon synchrotron emission. Contrary to models with equipartition ($a=1$) in relativistic particles, photo-meson production is not significant in proton dominated jets, since the synchrotron field is relatively weak (see Romero \& Vila 2008). 

The quoted luminosities correspond to the flux reported by AGILE and the upper X-ray limits, at a distance of the order of $\sim 2$ kpc. And accurate determination of the distance can be used to constrain the energy budget of the sources. At TeV energies the luminosities (due mainly to $pp$ interactions) are at the level of $10^{32}-10^{33}$ erg s$^{-1}$, and hence the sources are not detectable by current Cherenkov atmospheric telescopes, although they could fall within the sensitivity of upgraded instruments like HESS II and MAGIC II.

\section{Discussion and summary}

A crucial feature of a synchrotron-proton dominated jet is that a very strong magnetic field is necessary to produce detectable radiation in $\gamma$-rays. This means that the acceleration region should be located very close to the compact object (at $\sim 10^{8}$ cm in our models). Intrinsic absorption is then not very important since the low energy fields, which are responsible for the opacity to gamma-ray propagation, are weak. In models with a high content of primary electrons these effects, at the base of the jet, are very significant leading to a complete suppression of all emission above $\sim 100$ GeV (Romero \& Vila 2008). Such models produce a huge amount of X-rays, something that is not observed in the unidentified MeV-GeV sources.

In our model we introduce a hard injection spectrum in order to achieve a strong contrast between leptonic and hadronic peaks. A softer injection would reduce the $L_{\gamma}/L_{X}$ ratio. We notice that the losses in the high magnetic field strongly affect the overall leptonic particle spectrum. Observations with the LAT instrument of the Fermi satellite will allow us to determine the photon spectrum of these sources in the range 100 MeV $-$ 100 GeV, allowing a better determination of the physical parameters. 

We remark that IC models are possible for these type of sources only if the radiation is produced in a region with a significantly smaller ($\sim 10-100$ G) magnetic field, see for example Punsly et al. (2000). Such a field would occur at distances of $10^{12}-10^{13}$ cm from the compact object. How the injector of relativistic particles could be located there will be discussed elsewhere (see, nonetheless, Bosch-Ramon 2007). 

In summary, we propose a complete lepto/hadronic jet model to explain the unidentified variable AGILE sources. This model assumes a strong component of relativistic primary protons and takes into account all radiative processes that might occur at the base of the jets. The predicted SEDs are in accordance with what we know about of these sources. The jet model is independent of the nature of the donor star, so it could explain both low- and high-latitude galactic sources. Fermi observations will allow us to determine better constraints on the spectral features, then making possible inferences about the actual conditions in the sources. Development of models for $\gamma$-ray production far from the compact object and models that take into account external effects is in progress.              

\begin{acknowledgements}
We thank Mat\'{\i}as Reynoso for many fruitful discussions. GER acknowledges support by the Ministerio de Educaci\'on y Ciencia (Spain) under grant AYA 2007-68034-C03-01, FEDER funds. GSV thanks Nicol\'as Casco for his great help with numerical computations.
\end{acknowledgements}

\end{document}